\documentclass[rnote]{aa} 

\usepackage{graphicx}
\usepackage[varg]{txfonts}
\usepackage{natbib}
\bibpunct{(}{)}{;}{a}{}{,}

\begin{document}
\title{Particle energisation in a collapsing magnetic trap model: the relativistic regime}

\author{S.\ Eradat Oskoui \and T.\ Neukirch 
         }

\institute{School of Mathematics and Statistics, University of St Andrews, St. Andrews KY16 9SS,UK
              \\
              \email{se11@st-andrews.ac.uk} \\
               \email{tn3@st-andrews.ac.uk}
}

\date{Received March 26, 2014; accepted June 18, 2014}
 
\abstract{In solar flares, a large number of charged particles is accelerated to high energies. By which physical processes this is achieved is one of the
main open problems in solar physics. It has been suggested that during a flare, regions of the rapidly relaxing magnetic field can form a collapsing magnetic trap (CMT) and
that this trap may contribute to particle energisation.}
{ In this Research Note we focus on a particular analytical CMT model based on kinematic magnetohydrodynamics.
Previous investigations of particle acceleration for this CMT model focused on the non-relativistic energy regime. It is the specific aim of this Research Note
to extend the previous work to relativistic particle energies.}
{Particle orbits were calculated numerically using the relativistic guiding 
centre equations. We also calculated particle orbits using the non-relativistic guiding centre equations for comparison.}
{For mildly relativistic energies the relativistic and non-relativistic particle orbits mainly agree well, but clear 
deviations are seen for higher energies. In particular, the final particle energies obtained from the relativistic calculations are systematically lower than the
energies reached from the corresponding non-relativistic calculations, and the mirror points of the relativistic orbits are systematically higher than
for the corresponding non-relativistic orbits. }
{While the overall behaviour of particle orbits in CMTs does not differ qualitatively when using the relativistic guiding centre equations, there are a few systematic 
quantitative differences between relativistic and non-relativistic particle dynamics.}
{}


\titlerunning{Collapsing Magnetic Traps --The Relativistic Regime}

\maketitle
%

\section{Introduction}

Collapsing magnetic traps (CMTs) have been suggested as one of the mechanisms that might contribute to
particle energisation in solar flares \citep[e.g.][]{SomovKosugi}.
The basic idea behind CMTs is that charged particles will be trapped on the magnetic field lines below the 
reconnection region of a flare. 
The magnetic field will evolve into a state of lower energy, resulting in (a) a shortening of the field line length
and (b) an increase in the overall field strength.
Owing to the vast difference in length and time scales between the particle motion and the magnetic field evolution,
the particle motion can be described with very good accuracy by the guiding centre theory. The conservation of 
a particle's magnetic moment and the bounce invariant \citep[e.g.][]{Keith_grady_2012} give rise
to the possibility of increase in the particle's kinetic energy by betatron acceleration and by first-order Fermi acceleration \citep[e.g.][]{SomovKosugi,Boga_somov_2005}.
Several studies have focused on particle energisation and motion in CMT models with varying 
degrees of detail and have treated different aspects of the physical processes
\citep[e.g][]{Boga_Somov_2001,Kova_somov_2002,Somov_2003,Kar_kos_2004,Boga_somov_2005,Giul2005, Kar_Barta_2006, Boga_somov_2007,  grady:TN09, Boga_somov_2009, Mino_2010, Mino_2011,Keith_grady_2012,filatov:etal13,eradatoskoui:etal14}.

In this Research Note we used the relativistic guiding centre equations \citep[see e.g.][]{northrop} to extend to the relativistic regime the investigations 
of a specific analytic CMT model carried out 
with the non-relativistic guiding centre equations by 
\citet{Giul2005} 
and \citet{Keith_grady_2012}.
We compare particle motion and energisation in the
relativistic case with the non-relativistic 
case. 
In Sect. \ref{sec:model}  we outline of the CMT model of \citet{Giul2005} and 
introduce the relativistic guiding centre equations.
In Sect. \ref{sec:rel_vs_non_rel_cases} we present the results obtained from the relativistic 
particle orbit calculations by considering the effect of different initial conditions.
These relativistic results are
compared with results obtained from non-relativistic particle orbit calculations, in particular those reported by \citet{Keith_grady_2012}.
We present a summary of our results in Sect. \ref{sec:conclusions}

\section{Background}
\label{sec:model}

\citet{Giul2005} developed the basic theory for the construction of kinematic CMT models, and for brevity we refer the reader to their paper
for a detailed discussion 
We used the same coordinate system as in \citet{Giul2005} and \citet{Keith_grady_2012}:
the $x$ and $z$ coordinates run parallel to the solar surface, and
$y$ represents the height above the solar surface. We assumed that the $z$-components of the magnetic field and the velocity field vanish.
The magnetic field can then be described by a time-dependent flux function $A(x,y,t)$, with $\mathbf{B}(x,y,t) = \nabla A \times \mathbf{e}_z$.

A CMT model is then defined by choosing a form for the flux function $A_\infty(x,y)$ for time $t\to \infty$,
and by specifying a flow field $\mathbf{v}(x,y,t)$. Instead of defining the flow field
directly, it is given implicitly in the theory of \citet{Giul2005} by choosing a time-dependent transformation between Lagrangian and Eulerian coordinates, which 
facilitates solving the kinematic MHD equations.



We used the same magnetic field model and transformation as \citet{Giul2005}, 
\citet{Keith_grady_2012} and \citet{eradatoskoui:etal14}, that is,
\begin{eqnarray}
A_\infty &= & c_1\left[ \arctan\left(\frac{y_\infty+d_1}{x_\infty+w}   \right) - \arctan\left(\frac{y_\infty+d_2}{x_\infty-w}   \right)  \right],
\label{defA0}\\
  x_\infty &=& x,\label{x_infinity}\\
  y_\infty &=& (at)^{b} \ln\left[ 1 + \frac{y}{(at)^{b}}\right] \left\lbrace \frac{1+ \tanh\left[ (y - L_v  ) a_1\right]}{2}\right\rbrace \nonumber \\ 
            && + \left\lbrace \frac{1- \tanh\left[ (y - L_v ) a_1\right]}{2}\right\rbrace y  \label{y_infinity}.
 \end{eqnarray}
The parameter values are the same as in \citet{Giul2005}, \citet{Keith_grady_2012} and \citet{eradatoskoui:etal14}, namely
$w=0.5$,
 $d_1=d_2=1.0$, $a = 0.4$, $b = 1.0$, $L_v = 1.0$ and $a_1 = 0.9$. Here all length scale are normalised to $L=10^7$ m.
 The meaning of these parameters is explained in \citet{Giul2005}.


Because of the large difference between the time and length scales on which the electromagnetic fields of CMT model change compared with typical Larmor frequencies and radii of charged particle orbits,
the guiding centre approximation can be used to calculate particle orbits. We used the relativistic guiding centre equations as given by  \citet{northrop}:
\begin{eqnarray}
& \dot{\bf{R}}_{\perp}& = {{\frac{{\bf{b}}}{B(1 - {{{E}_{\perp}}^{2}/c^2B^{2}})}}} \times \left\lbrace {{- \left( 1-{\frac{{E_{\perp}}^2}{c^2B^2}} \right){\bf{E}}}} \right.\nonumber \\
&+& \left. \frac{\mu_{r}}{\gamma q} \nabla \left[ B\left( 1-{\frac{{E_{\perp}}^2}{c^2B^2}}\right)^{1/2} \right]  + \frac{m \gamma }{q}\left( v_{\parallel} \frac{d{\bf{b}}}{dt} + \frac{d{\bf{u}}_{E}}{dt} \right)  \right. \nonumber \\
&+& \left. \frac{v_{\parallel}E_{\parallel}}{c^{2}}{\bf{u}_{E}} + \frac{\mu_{r}}{\gamma q}\frac{{\bf{u}}_{E}}{c^2}\frac{\partial}{\partial t}\left[ 
B\left( 1-{\frac{{E_{\perp}}^2}{c^2B^2}}\right)^{1/2}
\right] 
\right\rbrace, \label{dvperpdt}
\end{eqnarray}
\begin{eqnarray}
m\frac{ d(\gamma {v}_{\parallel})}{d t} = \frac{d{p}_{\parallel}}{dt} = {q{E}_{\parallel}} &-& {{\frac{\mu_{r}}{\gamma} \frac{\partial}{\partial s}\left[ B\left( 1-\frac{{E_{\perp}}^2}{c^2B^2}
\right)^{1/2} 
\right]}}  \nonumber \\
& +& {{m \gamma {\bf{u}}_{E} \cdot (\frac{\partial {\bf{b}}}{\partial t} +   \frac{{v}_{\parallel}}{\gamma}\frac{\partial {\bf{b}}}{\partial s}  + {\bf{u}}_{E} \cdot \nabla{\bf{b}})}},
 \label{dvpardt}
\end{eqnarray}
\begin{eqnarray}
 \frac{d \gamma}{dt} = \frac{q}{m c^2}(\dot{\bf{R}}_{\perp} + v_{\parallel}){\bf{b}}\cdot{\bf{E}} + \frac{\mu_{r}}{mc^2 \gamma} \frac{\partial B}{\partial t}. \label{eqn_dgammadt}
\end{eqnarray}
Here,
$m$, $q$, and $c$ are the particle rest mass, particle charge, and the speed of light, respectively,  $\mathbf{R}$ denotes the guiding centre position, 
$\mathbf{E}$ and $\mathbf{B}$ the electric and magnetic fields at the guiding centre position, $\dot{\mathbf{R}}_\perp$ is the guiding centre velocity perpendicular to the magnetic field direction, and
$v_\parallel$ is the guiding centre velocity parallel to the magnetic field direction.
$ \mathbf{b}$ denotes the unit vector along magnetic field lines, 
$s$ is the coordinate (arc length) along magnetic field lines, 
${\mathbf{u}}_E = \mathbf{E} \times \mathbf{b}/B$ is the ${\mathbf{E}} \times {\mathbf{B}}$ drift velocity, and $\mu_r = m  ({\gamma v_\perp})^2/2B$ is the relativistic magnetic moment, with $v_\perp$ the 
perpendicular particle velocity associated with its gyration around the magnetic field lines.

In practice, the CMT model provides the expressions for the electric and the magnetic field and their spatial and temporal derivatives.
The guiding centre equations (\ref{dvperpdt}) - (\ref{eqn_dgammadt}) are then used to calculate the particle trajectories in the CMT model. 
Because the electric field in our CMT model is derived from the ideal Ohm's law it only has a component perpendicular to the magnetic field, and 
$E_\perp \ll cB$ because the flow velocities in the CMT model are well below the speed of light. We do not state the non-relativistic guiding centre equations here \citep[see e.g][]{northrop,Giul2005,Keith_grady_2012},
but they can in principle be recovered from the relativistic equations (\ref{dvperpdt}) and (\ref{dvpardt}) letting $\gamma \to 1$ and $ (E_\perp/cB)^2 \to 0$ in expressions that contain this term.

To discuss the energy gain experienced by particles in the CMT we used the following expressions for the kinetic energy:
in the relativistic regime, the kinetic energy of the particle is given by
\begin{eqnarray}
 E_k = m c^2 (\gamma - 1) \label{rel_E_k}.
\end{eqnarray}
The rest mass energy of electrons is $mc^2 \approx 511\mbox{keV}$.
We recall that ${v_{\perp}}^2 = 2 \mu_r B/(m \gamma^2)$;
therefore  gyro-averaged Lorentz factor can be expressed as
\begin{eqnarray}
 \gamma(t) = \sqrt{\frac{1 + [2 \mu_r B(t)]/c^2}{1- [v_{\parallel}^2(t) + u_{E}^2(t)]/c^2}}.
\end{eqnarray}
The Lorentz factor changes in time because the electromagnetic fields and $v_\parallel$ are time dependent. As we describe below, the Lorentz factor increases on average with time in a CMT.

For comparison, the non-relativistic kinetic energy $E_K$ is given by 
\begin{eqnarray}
 E_k = \frac{m{v_\parallel}^2}{2} + \mu B + \frac{m {u_E}^2}{2} \label{non_E_k},
\end{eqnarray}
where $m{v_\parallel}^2/2$ is the parallel energy of the particle, $\mu B$ is the energy associated with the gyrational motion, and $m {u_E}^2/2$ is the energy associated with the main
perpendicular drift of the guiding centre, which is usually very small compared to the other terms \citep[see e.g.][]{Keith_grady_2012}. 


\section{Particle trajectories and energy gain} 

\label{sec:rel_vs_non_rel_cases}


We first investigated a typical particle orbit and the evolution of its kinetic 
energy. We chose an orbit starting at the initial position $(x, y) = (0,4.2)$, an initial 
pitch angle $160.4^{\circ}$, and an initial energy $5.5$ keV. 
These initial condition were chosen because they are very similar to one of the examples discussed in \citet{Keith_grady_2012}.
We then compared this with an orbit with the same initial position and pitch angle, but with an an already mildy relativistic initial energy of $200$ keV.
\begin{figure*}
\centering
\includegraphics[width=0.4\hsize]{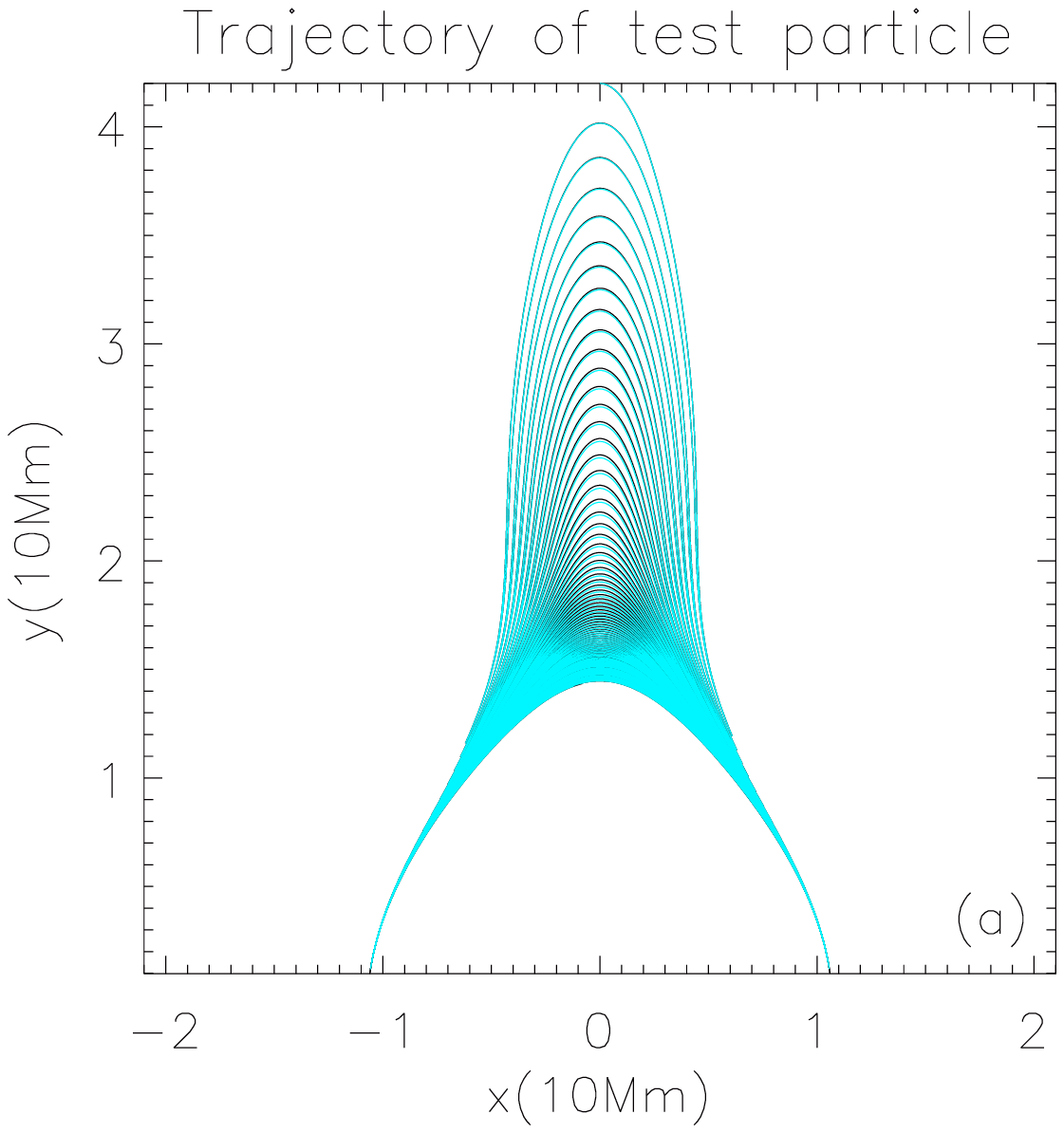}
\includegraphics[width=0.4\hsize]{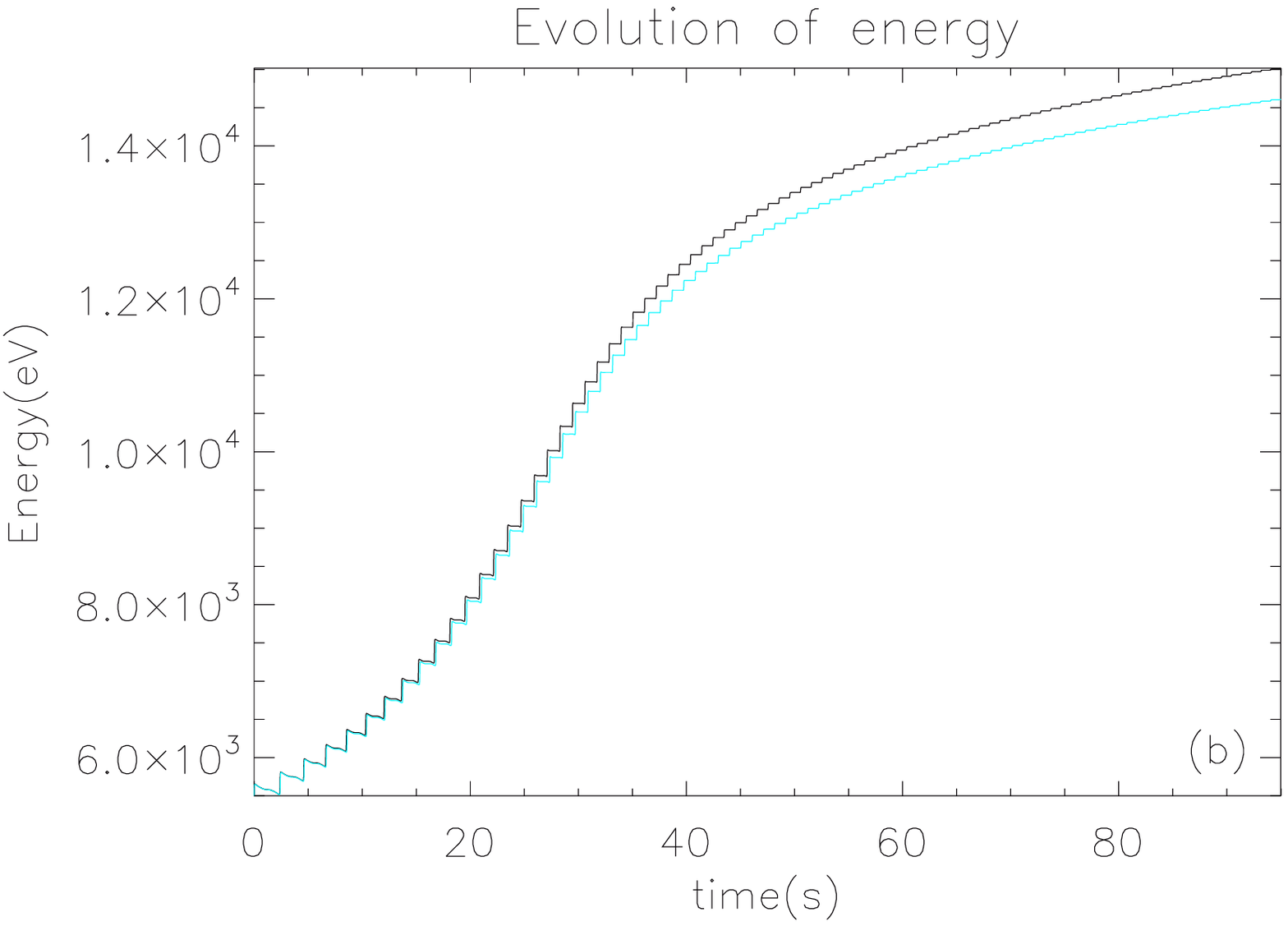}

\includegraphics[width=0.4\hsize]{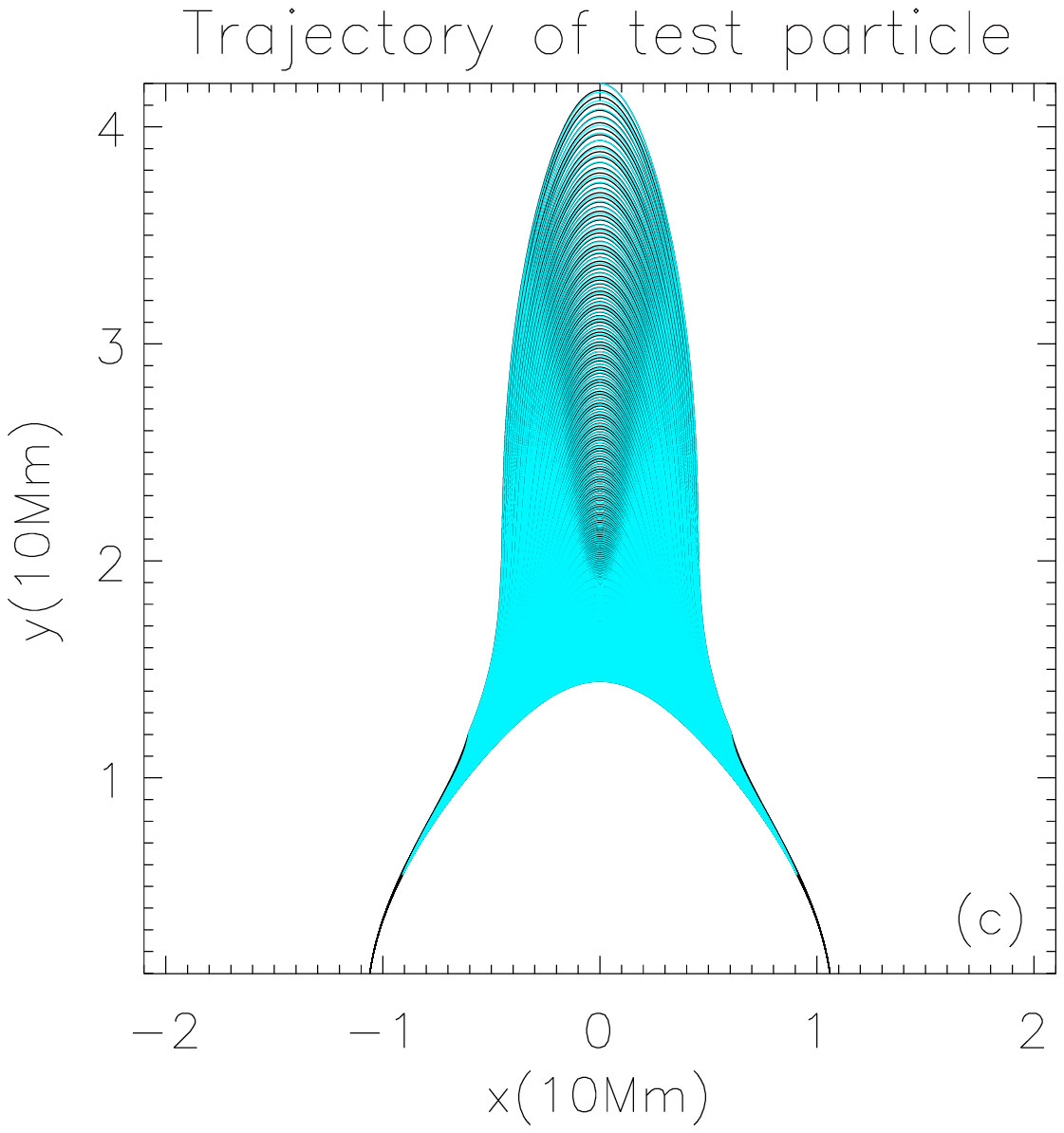}
\includegraphics[width=0.4\hsize]{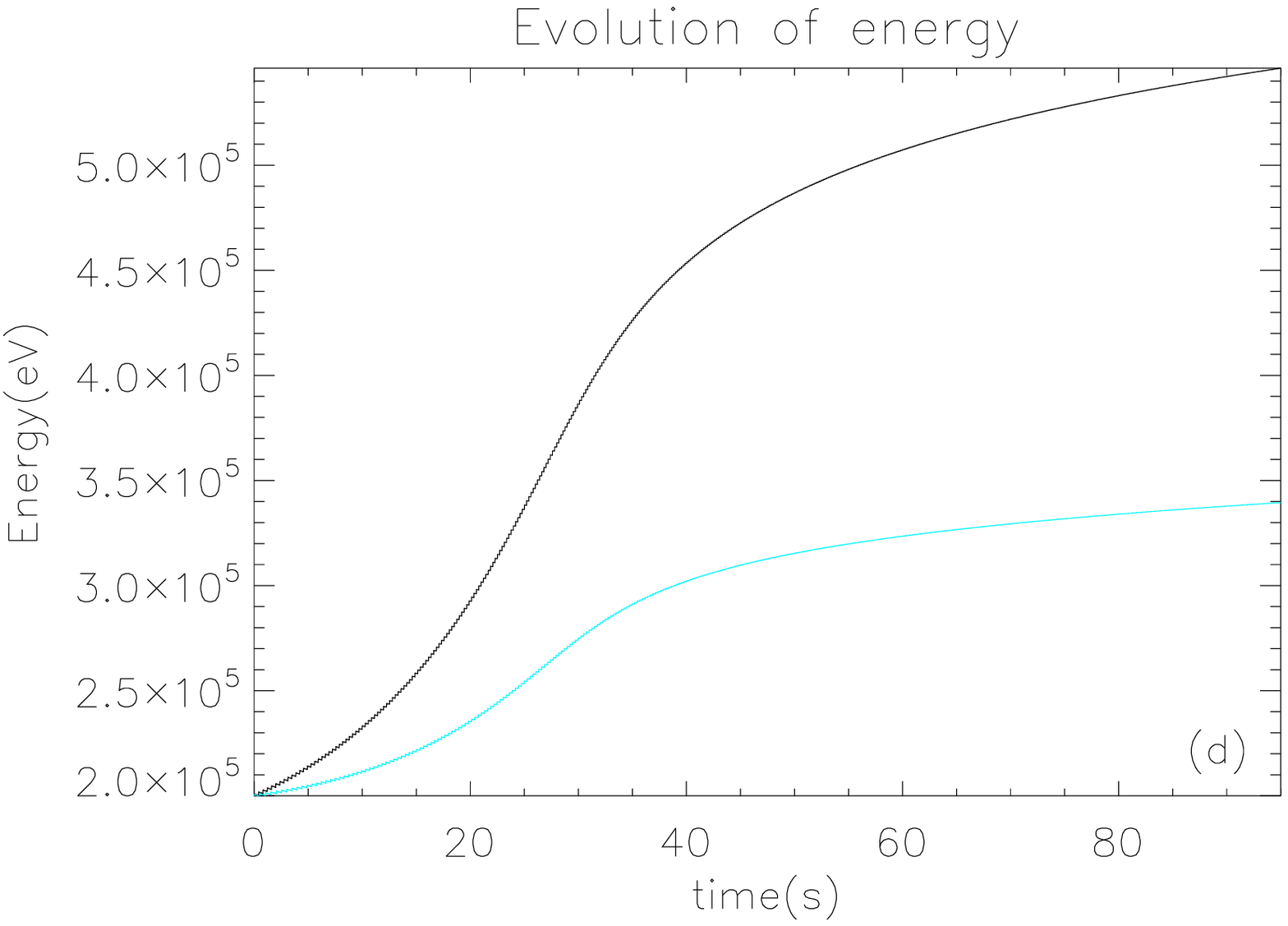}
   \caption{
  Particle orbits and energy evolution for test particle orbits starting at $(x,y)=(0,4.2)$. 
 The turquoise curves represent the relativistic calculations, the black curves represent the 
non-relativistic calculation.
 Panels (a) and (c) show the particle trajectories, panels (b) and (d) the energy evolution 
for initial energies $5.5$ keV (panels (a) and (b)) and $200$ keV (panels (c) and (d)). 
}
  \label{Fig:orbit_and_energy}
\end{figure*}

The results of the particle orbit calculations are shown in Fig.~\ref{Fig:orbit_and_energy}.
As previously discussed by, for example, \citet{Keith_grady_2012} and
\citet{eradatoskoui:etal14}, the $\mathbf{E}\times \mathbf{B}$-drift is by far the dominant drift in this CMT model, and thus particles remain on the same field line at all times. This explains the spatial overlap of the
relativistic and non-relativistic trajectories. Note, however, that this does not necessarily imply that the non-relativistic and relativistic trajectory are always at the same position at the same time.

Panel (b) in Fig.~\ref{Fig:orbit_and_energy} shows the time evolution of the kinetic energy.
%
%
%
Obviously, both graphs start at $5.5$ keV, but then begin to move apart. The final shown energy of the non-relativistic orbit 
is about $15.02$ keV, whereas the final energy for the relativistic case is about $14.61$ keV, giving a difference of about $0.41$ keV
or $2.7\%$. The particle velocity at the final time shown is roughly $0.2$ c. 

As is to be expected the relativistic kinetic energy curve lies systematically below the non-relativistic kinetic energy curve.
Another way to express this fact would be to say that in the relativistic case it takes a longer time to reach the same energy than
in the non-relativistic case. 
In both cases, one can see the step-like pattern in the energy curves, which is particularly prominent in the initial phases. 
In the non-relativistic case, this pattern
has been associated with changes in the parallel energy due to the curvature-related terms in the equation for the parallel velocity
\citep[e.g][]{Giul2005,Keith_grady_2012}. In the relativistic case, it is difficult to split the kinetic energy into parallel and perpendicular components,
but the process seems to operate in the same way.


The case with a starting energy of $5.5$ keV shows only very weak relativistic effects. However,
when the initial energy is increased to the already 
mildly relativistic value of $200$ keV, the discrepancies between the relativistic calculation and the non-relativistic calculation
become much more pronounced. We would like to emphasise that the non-relativistic calculation was made only for comparison purposes.
With an initial value of $\gamma = 1.39$ giving $v = 0.695$ c, it is clear that only a fully relativistic particle orbit calculation can yield physically
correct results.

We show the results for this orbit in the lower two panels of Fig. \ref{Fig:orbit_and_energy} (panels (c) and (d)).
For this case, even changes in the particle orbits are visible (Fig. \ref{Fig:orbit_and_energy} (c)). The relativistic orbit has systematically higher mirror points than the non-relativistic orbit. 
We also found this behaviour for other orbits and discuss the reasons for this below.

Since the initial kinetic energy 
$E_{init}$ is now already mildly relativistic, a much larger difference between the energy graphs can be seen 
in panel (d) of Fig. \ref{Fig:orbit_and_energy}. At the final time of the calculation, the 
kinetic energy in the non-relativistic case is about $546$ keV, whereas the kinetic energy of the 
relativistic case it is about $339$ keV. This is a discrepancy of about $207$ keV or roughly $38\%$. This is a 
significant difference in energy compared with the case with $E_{init} = 5.5$ keV. The final Lorentz 
factor and the particle velocity for the relativistic case 
are $\gamma = 1.664$ and $v = 0.799$ c, which again corroborates the need for a relativistic calculation.

From our case studies, we found two main differences between the relativistic and non-relativistic orbit calculations. 
Firstly, the final particle energy calculated using the relativistic equations is always lower than the non-relativistic case. 
This is easily explained because the 
Lorentz factor grows nonlinearly when the particle velocity approaches c, which diminishes the energy gain in the relativistic regime.
Secondly, we found that the mirror points in the relativistic 
cases are systematically higher than the non-relativistic cases. At first sight, this seems counterintuitive because the mirror term
in Eq. (\ref{dvpardt}) scales with $\gamma^{-1}$ and therefore one would expect mirror points to be lower for higher values of $\gamma$.
However, a careful investigation of the terms in Eq. (\ref{dvpardt}) reveals that the terms associated with field line curvature also
decrease with increasing $\gamma$. These terms are responsible for increasing the parallel velocity component, and if this
process becomes less efficient with increasing $\gamma$, the mirror points are expected to be higher in the CMT than
for the corresponding non-relativistic case.
%

\begin{figure}
\resizebox{\hsize}{!}{
\includegraphics{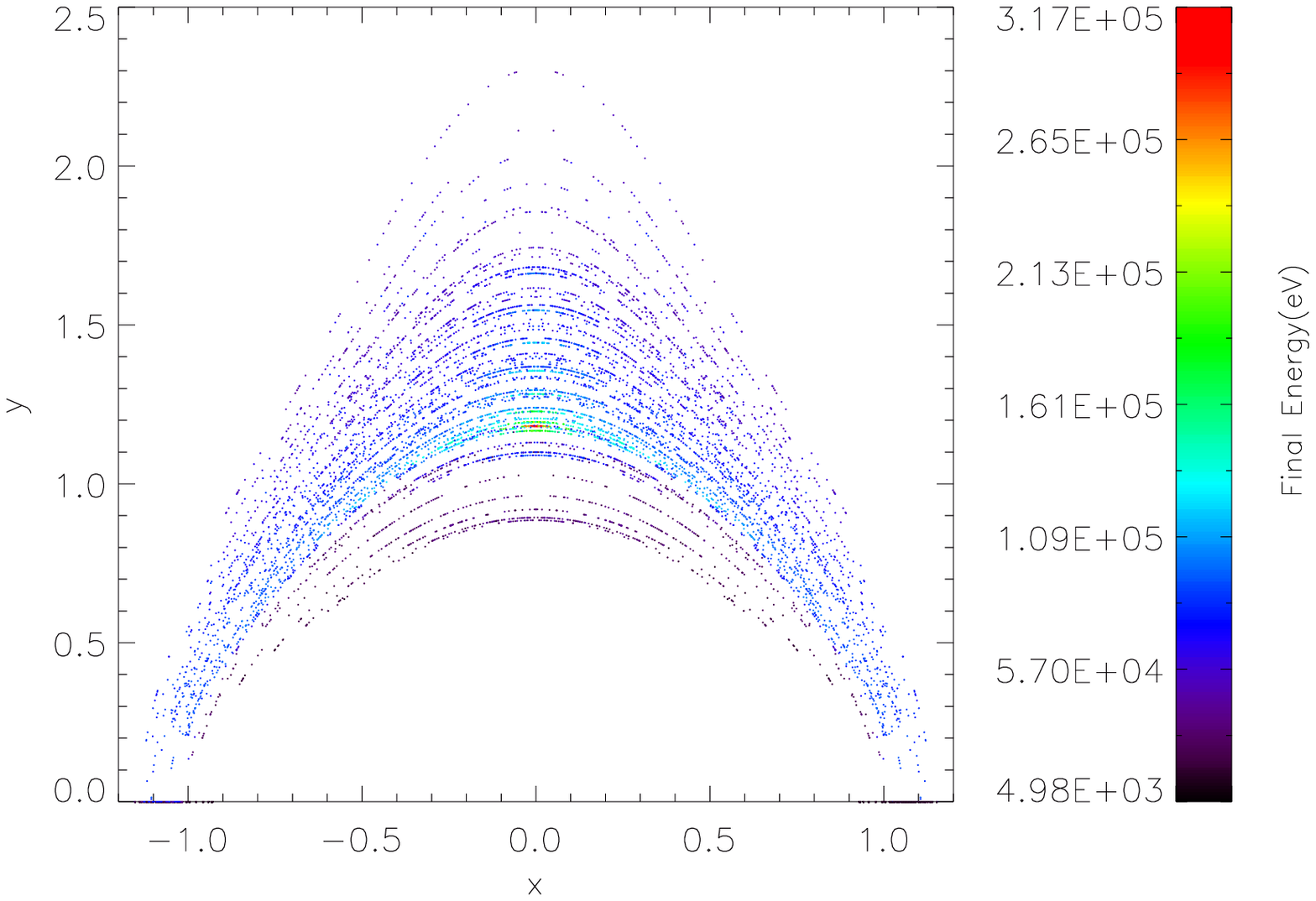}}
\resizebox{\hsize}{!}{
\includegraphics{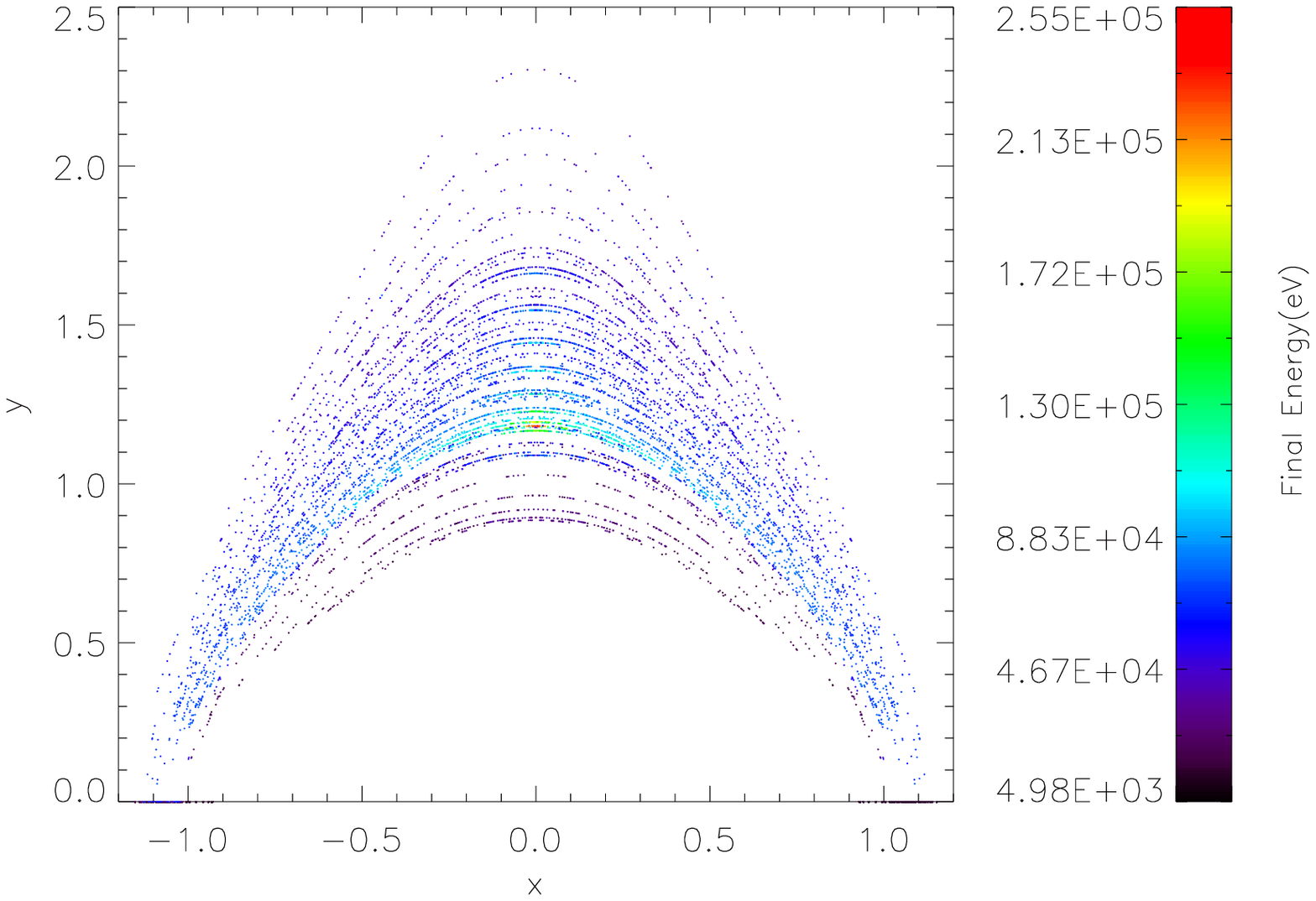}}
  \caption{Snapshot of the particle position at the final time of the calculation. 
  The colour code indicates the kinetic energy at that time. \bf Top: the non-relativistic case; bottom: the relativistic case. 
}
  \label{Figlfinalpos}
\end{figure}

Our previous findings were based on the single case studies presented above. However, we calculated particle orbits for a whole range of initial conditions,
similar to the non-relativistic investigation by \citet{Keith_grady_2012}. The conclusion from the case studies are qualitatively similar, although
there are of course quantitative differences depending on initial conditions.
A key result found by \citet{Keith_grady_2012} was that the particle orbits that gain the most energy during the 
collapse of the trap have initial pitch 
angles close to $90^{\circ}$ and remain trapped at the field line apex (loop top). The particles
with the highest energy gain have initial positions in a weak magnetic field region in the middle of the trap. 
We found the same result when we used the same set of initial conditions for the relativistic case. 
For comparison, we show in Fig. \ref{Figlfinalpos} the positions of a number of particle orbits at the final time of the
calculation with the colour code indicating their kinetic energy at that time. 
The range of final energies in the non-relativistic case is approximately $4.98 - 317$ keV, whereas it is $4.98-255$ keV, in the relativistic case,
 consistent with our previous findings that lower energies are reached in the relativistic case.


\section{Summary and conclusions} 

\label{sec:conclusions}

In this Resarch Note we have extended previous investigations of particle energisation in the 
CMT model of \citet{Giul2005} to the relativistic regime, using the relativistic  guiding centre equations.
%
%
Qualitatively, the particle orbits obtained using the relativistic guiding centre equations are very similar to the results
obtained by \citet{Keith_grady_2012} with the non-relativistic guiding centre equations, in particular regarding the trapping of the 
particle orbits with the strongest increase in energy in the centre of the CMT.
%
We found two main differences between the relativistic and non-relativistic results, in particular for larger initial energies.


The final particle energy calculated using the 
relativistic approximation is always lower than in the non-relativistic case. In general, 
the difference in final energy grows with increasing initial energy.
This is due to the Lorentz factor, which reduces the size of the terms in the 
equation of motion that contribute to energy gain. It should be borne in mind, however, that the non-relativistic 
equations cease to be accurate for higher energies and were presented here only for comparison.


In the relativistic case the mirror points of the particle orbits are 
located systematically higher than in the corresponding the non-relativistic case. 
We explained this by investigating the size of the terms that are related to curvature in the equation for
the parallel velocity, which decrease with increasing $\gamma$ factor. This seems to have a stronger effect on the mirror point location
than the change of the actual mirror term, which also decreases with increasing $\gamma$ factor.



\begin{acknowledgements}
The authors acknowledge financial support by the UK's Science and Technology Facilities Council through a Doctoral
Training Grant (SEO) and Consolidated Grant ST/K000950/1 (SEO and TN).

\end{acknowledgements}

\bibliographystyle{aa}
\bibliography{paper}

\begin{thebibliography}{17}
\expandafter\ifx\csname natexlab\endcsname\relax\def\natexlab#1{#1}\fi

\bibitem[{{Bogachev} \& {Somov}(2001)}]{Boga_Somov_2001}
{Bogachev}, S.~A. \& {Somov}, B.~V. 2001, Astronomy Reports, 45, 157

\bibitem[{{Bogachev} \& {Somov}(2005)}]{Boga_somov_2005}
{Bogachev}, S.~A. \& {Somov}, B.~V. 2005, Astronomy Letters, 31, 537

\bibitem[{{Bogachev} \& {Somov}(2007)}]{Boga_somov_2007}
{Bogachev}, S.~A. \& {Somov}, B.~V. 2007, Astronomy Letters, 33, 54

\bibitem[{{Bogachev} \& {Somov}(2009)}]{Boga_somov_2009}
{Bogachev}, S.~A. \& {Somov}, B.~V. 2009, Astronomy Letters, 35, 57

\bibitem[{{Eradat Oskoui} {et~al.}(2014){Eradat Oskoui}, {Neukirch}, \&
  {Grady}}]{eradatoskoui:etal14}
{Eradat Oskoui}, S., {Neukirch}, T., \& {Grady}, K.~J. 2014, \aap, 563, A73

\bibitem[{{Filatov} {et~al.}(2013){Filatov}, {Melnikov}, \&
  {Gorbikov}}]{filatov:etal13}
{Filatov}, L.~V., {Melnikov}, V.~F., \& {Gorbikov}, S.~P. 2013, Geomagnetism
  and Aeronomy, 53, 1007

\bibitem[{{Giuliani} {et~al.}(2005){Giuliani}, {Neukirch}, \&
  {Wood}}]{Giul2005}
{Giuliani}, P., {Neukirch}, T., \& {Wood}, P. 2005, \apj, 635, 636

\bibitem[{{Grady} \& {Neukirch}(2009)}]{grady:TN09}
{Grady}, K.~J. \& {Neukirch}, T. 2009, \aap, 508, 1461

\bibitem[{{Grady} {et~al.}(2012){Grady}, {Neukirch}, \&
  {Giuliani}}]{Keith_grady_2012}
{Grady}, K.~J., {Neukirch}, T., \& {Giuliani}, P. 2012, \aap, 546, A85

\bibitem[{{Karlick{\'y}} \& {B{\'a}rta}(2006)}]{Kar_Barta_2006}
{Karlick{\'y}}, M. \& {B{\'a}rta}, M. 2006, \apj, 647, 1472

\bibitem[{{Karlick{\'y}} \& {Kosugi}(2004)}]{Kar_kos_2004}
{Karlick{\'y}}, M. \& {Kosugi}, T. 2004, \aap, 419, 1159

\bibitem[{{Kovalev} \& {Somov}(2002)}]{Kova_somov_2002}
{Kovalev}, V.~A. \& {Somov}, B.~V. 2002, Astronomy Letters, 28, 488

\bibitem[{{Minoshima} {et~al.}(2010){Minoshima}, {Masuda}, \&
  {Miyoshi}}]{Mino_2010}
{Minoshima}, T., {Masuda}, S., \& {Miyoshi}, Y. 2010, \apj, 714, 332

\bibitem[{{Minoshima} {et~al.}(2011){Minoshima}, {Masuda}, {Miyoshi}, \&
  {Kusano}}]{Mino_2011}
{Minoshima}, T., {Masuda}, S., {Miyoshi}, Y., \& {Kusano}, K. 2011, \apj, 732,
  111

\bibitem[{{Northrop}(1963)}]{northrop}
{Northrop}, T.~G. 1963, {The Adiabatic Motion of Charged Particels} (Wiley, New
  York)

\bibitem[{{Somov} \& {Bogachev}(2003)}]{Somov_2003}
{Somov}, B.~V. \& {Bogachev}, S.~A. 2003, Astronomy Letters, 29, 621

\bibitem[{{Somov} \& {Kosugi}(1997)}]{SomovKosugi}
{Somov}, B.~V. \& {Kosugi}, T. 1997, \apj, 485, 859

\end{thebibliography}

\end{document}